\documentclass[journal]{IEEEtran}

\ifCLASSINFOpdf
\usepackage[pdftex]{graphicx,color}
\else
\fi
\usepackage[cmex10]{amsmath}
\usepackage{amssymb}
\usepackage{mathcomp} 
\newcommand{\red}{\color{black}}
\newcommand{\ver}{\color{black}}

\begin{document}

\title{Trigger-disabling Acquisition System for Quantum Key Distribution failsafe against Self-blinding}

\author{M. Bawaj\(^1\), M. Lucamarini\(^1\), R. Natali\(^{1}\), G. Di Giuseppe\(^{1, 2}\), and P. Tombesi\(^{1, 2}\) \\
\(^1\) \hspace{1pt} School of Science and Technology,~Physics
Division,~University of Camerino, I-62032 Camerino (MC), Italy\\
\(^2\) \hspace{1pt} CriptoCam S.r.l., via Madonna delle Carceri 9, I-62032 Camerino (MC), Italy\\
\thanks{Corresponding author email: mateusz.bawaj@unicam.it.}
\vspace{12pt} \today }
%
%
\maketitle

\begin{abstract}
Modern single-photon detectors based on avalanche photodiodes
offer increasingly higher triggering speeds, thus fostering their
use in several fields, prominently in the recent area of Quantum
Key Distribution.
{\red It is well known that after a detection event these
detectors loose their single photons sensitivity for a period of
time generically called \textit{dead time}.}
%
%
If the acquisition system connected to the detector is not
properly designed, {\red and the triggering function not properly
controlled,} efficiency issues {\red can} arise when the
triggering rate is faster than the inverse of detector's
dead-time. Moreover, when this happens with two or more detectors
used in coincidence, a security risk called ``self-blinding'' can
jeopardize the
distribution of a secret quantum key.\\
In this paper we introduce a trigger-disabling circuitry based on
an FPGA-driven feedback loop, so to avoid the above-mentioned
inconveniences. In the regime of single-photon-attenuated light,
the electronics dynamically accept a trigger only after detectors'
complete recovery from dead-time. This technique proves useful to
work with detectors at their maximum speed and to increase the
security of a quantum key distribution setup.
\end{abstract}


\section{Introduction}

\IEEEPARstart{I}{n order} to detect a light signal attenuated at
the single-photon level~\cite{note1}, the Single-Photon Avalanche
Diode (SPAD) is of widespread use. SPADs have been employed to
detect stars feeble light~\cite{ZTC+07}, to perform a sharp
optical time-domain reflectometry~\cite{ELZ+10} and in the vast
majority of Quantum Key Distribution (QKD) realizations reported
thus far~\cite{BB84,GRT+02,Scarani2009}, both in optical fibres
and in free space. In particular, fibre-based QKD at the
wavelength of $1550$~nm, is an emerging technology which promises
a high security level in telecommunications while retaining quite
a high transmission {\red rate}. 
In this wavelength range, the SPAD features quantum efficiencies
up to $25\%$~\cite{idq} and trigger rates of more than
1~GHz~\cite{Namekata2006,Yuan2007,Dixon2009,Zhang2009,Namekata2009}.

{\red Let us briefly describe some}
particular aspects of the SPAD. First, in order to {\red increase
the detector's gain, the detector is operated in the so called
``Geiger mode'', where a reverse bias voltage higher than the
detector's breakdown voltage is applied. If, despite the high
gain, the dark count rate is low, the SPAD can be operated in
``free running'' mode; this usually happens at wavelengths close
to the visible range, e.g. 800 nm. If the dark count rate is high,
as it happens in the infrared domain, e.g. at wavelengths of 1550
nm, then}
%
%
the SPAD is usually 
{\red run in ``gated mode'', where the bias voltage is raised
above the breakdown voltage only for a short period of time (the
gate), when the photon to be detected is expected to arrive.}
This requires that the SPAD is well synchronized with the light
source and the rest of the acquisition system.
{\red Another important point to consider is that the dark count
rate can be increased by a non zero afterpulse probability: a
detection event is given in the SPAD by an avalanche of electric
charges, some of which can remain trapped in the junction and can
be released in the following gates, thus giving rise to additional
counts not corresponding to the arrival of new photons.}
To reduce the afterpulse probability, an {\red afterpulse
blocking} electronics{\red ~\cite{MBH04}} is usually added to the
SPAD with the aim of ``freezing'' it after the emission of an
avalanche, for a time interval decided in advance by the operator.
For example, one of the most popular single-photon detector in the
third Telecom window, the id201 from IDQ~\cite{idq}, {\red
features an afterpulse probability 
going to nearly zero in}
about 4~$\mu$s.~\cite{note2}. This entails that, in order to
minimize dark counts, a {\red blocking} time of at least 10~$\mu$s
should be applied to this detector.
%
%
During the {\red blocking time,
the detector is not capable to detect single photons anymore.
Hereafter, we generically define ``dead time'' any interval of
time in which the SPAD looses its single-photon sensitivity. This
can be due either to the natural response of the detector or to
the afterpulse blocking circuit.
Moreover, we focus on a SPAD run in gated mode, suitable for
working in a high-noise regime, leaving the free-running mode
analysis for a future work.
%
}

In order to use a {\red gated-mode} SPAD, it is necessary to
prepare an acquisition system capable of sending a trigger to it,
read its output in a synchronized way and write the result into a
memory location. However, during the preparation of such a system,
we realized that a few problems arise when the triggering rate of
the acquisition system is higher than the inverse of detector's
dead-time. This situation is much more widespread than commonly
believed. In fact, the acquisition system is usually much faster
than the SPAD and its triggering rate coincides with the maximum
rate allowed by the detector. For instance, the same id201
mentioned above, accepts a maximum trigger rate of 8~MHz, which is
much higher than the 100~KHz corresponding to the inverse of the
dead-time value given above, i.e. 10~$\mu$s.

One problem related to a SPAD running at its maximum trigger rate
and with a {\red non-zero} dead-time, 
is the readout. If the acquisition system tries to read the SPAD
{\red during the dead-time period}
it only collects a sequence of futile {\red counts} 
which do not correspond to true detection events.
The removal of these {\red counts} 
by a postprocessing algorithm is simple but very inefficient,
time- and memory-consuming. A second, more important, problem is
that when two or more SPADs are present in the same detection
apparatus, one detector can be active while the other is blind.
This happens, e.g., when one of the two detectors registers a
photon and enters the dead-time period while the other does not
register any photon and remains active. This situation can be
exploited to control the detectors of the receiving unit, as
described in{\red ~\cite{Weier11}.
Furthermore, in a wholly passive way, this flaw can be exploited
to threaten the security of the QKD technology, by making the key
distilled by the users not perfectly random and not entirely
secret, as we shall show.}

We {\red group these security breaches under the name of}
``self-blinding'', because {\red they are} 
caused {\red mainly} by an improper use of the equipment by the
legitimate users.

In this paper we describe an acquisition system that allows for a
proper triggering of single-photon detectors. We use an FPGA-based
electronics to disable not only the trigger driving the SPAD gates
but also the one driving the readout and the write up of the data
to the memory. Hence useless data are no more registered by the
acquisition system; the memory is better organized and the
postprocessing procedures require a shorter time. Moreover the
triggering technique can be easily extended to two or more
detectors, so to avoid { self-blinding}.

The paper is organized as follows. In
Section~\ref{sec:real_time_trig} we describe our technique and
apply it to a single SPAD. In Section~\ref{sec:pair-det-risk-QKD}
we consider the situation with a pair of SPADs and describe the {
security risks connected to self-blinding}. In
Section~\ref{sec:exp-coinc} we apply our technique to a { real
setup} and demonstrate its effectiveness.
Section~\ref{sec:conclusion} is left for concluding remarks.

\section{Real-time disabling of the trigger}
\label{sec:real_time_trig}

\begin{figure}[!h]
\centering
\includegraphics[width=0.9\columnwidth]{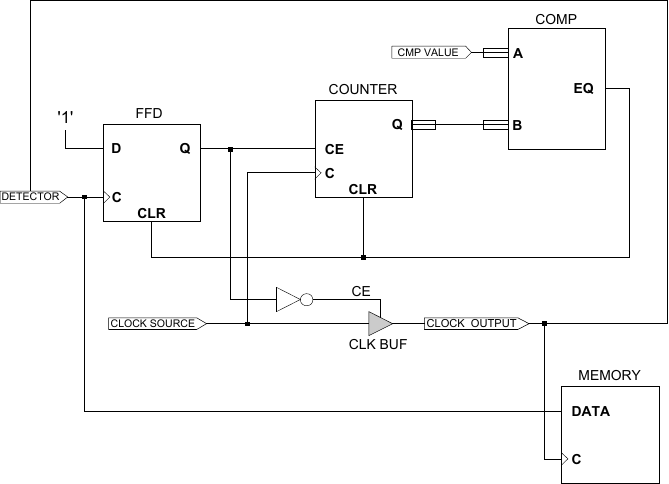}
\caption{Conceptual scheme of clock manager. FFD - D flip-flop latch, COUNTER - binary counter, COMP - binary comparator, CLK BUF - clock buffer, MEMORY - cache memory for storing detection results. For clarity the scheme does not include time correlations. To synchronize recording time and detection event additional elements are required.} \label{fig_scheme}
\end{figure}

The trigger-disabling acquisition system is schematically shown in
Figure~\ref{fig_scheme}.
The main element of the circuit is the clock buffer (CLK BUF)
with clock enable CE input (bottom part of the figure). Its
specifications are given in~\cite{XilinxUG:Clocking}. This buffer
drives the clock output which in turn is connected to the
detector's trigger input and memory clock input.
In normal regime, the D~flip-flop (FFD) is reset; the output is
inverted and then fed into the clock buffer so the trigger clock
is simply transferred from the input to the output and then to the
detector. When a positive detection occurs, the rising edge of the
SPAD's avalanche drives the change in the FFD output, { thus
disabling the main clock and enabling the COUNTER}. The SPAD and
the MEMORY will not see any trigger in this state. The FFD remains
in this state until it is reset. The asynchronous reset occurs
after a given number of main clock cycles, depending on the
COUNTER and the binary comparator (COMP) level indicated as
CMP~VALUE in Fig.~\ref{fig_scheme}. All the circuitry is realized
in an FPGA board (Xilinx SP605).

The trigger-disabling circuit has a main constraint  {i.e. the}
response time-interval $\tau_{resp}$ that goes from the SPAD
avalanche to the CLK BUF must be smaller than the inverse of the
trigger rate $\omega_{trig}$:
\begin{equation}\label{eq:phys-cons}
    \tau_{resp} < \frac{1}{\omega_{trig}}.
\end{equation}
If this condition is not fulfilled, the trigger pulses continue to
arrive at the detector until the clock buffer succeeds in
disabling the trigger. This would result in maximal frequency
limitation or in one or more futile zeroes registration by the QKD
apparatus's cache memory.

Several factors contribute to the total response time of the
feedback loop: the detector response (28~ns for
id200/id201~\cite{idq}, which are the SPADs used in our setup),
the length of cables (8~ns) and circuit inside FPGA (2~ns). This
sets a maximum limit for the trigger-disabling technique with our
current electronics at 26~MHz, which is much above the maximal
trigger rate for id201. Shortening the total response time would
increase the maximal frequency of the circuit. However, among the
above delays, only the one caused by cables can be decreased
easily without changing any other piece of hardware.

\begin{figure}[!h]
\centering
\includegraphics[width=3in]{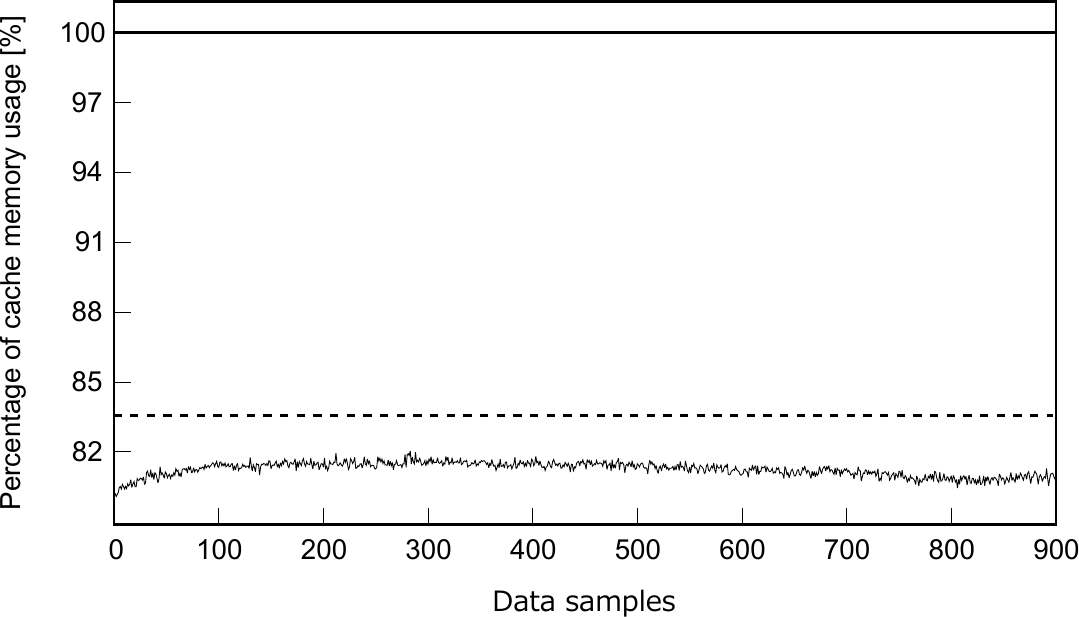}
\caption{Number of useful triggers per data sample. The lower
solid curve shows percentage of useful triggers in cache in case
of a standard triggering technique. It lays around a percentage
value of 81\%. When trigger-disabling technique is applied all the
triggers are useful, thus reaching a percentage of 100\% (upper
solid line). The dotted line is a theoretical prediction when
$\mu=0.1$ and $\eta = 0.1$, providing a percentage of {\ver
83.4\%}. Nine hundred samples were taken in 12
hours~\cite{note4}.}
\label{fig:triggers}
\end{figure}

In order to demonstrate the circuit working, we { applied our
electronics to a real setup, composed by a laser source (PicoQuant
LDH-P-1550) triggered at 4~MHz, an optical attenuator, which fixes
the average photon number at $\mu \simeq 0.1$, and a SPAD id201
set at 2.5~ns gate and 10\% efficiency. In this configuration, we
analyzed} the percentage of cache memory occupied by useful
triggers over a total duration of 12 hours~\cite{note4}. The
results are plotted in Fig.~\ref{fig:triggers}.
If trigger-disabling is applied, the cache memory is occupied by
useful data only, entailing that all the triggers sent to the SPAD
have been effectively used for a meaningful acquisition. By
consequence, the trigger-percentage per cache reaches the maximum,
100\%. On the contrary, if the disabling technique is not applied,
the total amount of useful counts per cache varies according to
the photon emission probability and detector dead-time. In
particular, a lower percentage corresponds to a situation with
more counts by the SPAD, i.e. to a higher intensity arriving at
the detector. In fact, the more counts at the SPAD, the higher the
number of futile zeroes present in the cache memory. Any futile
zero must be removed from the cache which causes that the
percentage of useful triggers remains significantly lower than
100\%. There are two ways of removal, either by post-processing or
by disabling the trigger before it arrives to the SPAD. In any
case the post-processing consumes more time than a hardware
solution~\cite{note4}.

Further performance analysis shows that the amount of removed
triggers from cache depends on the product of the average photon
number $\mu$ and detector quantum efficiency $\eta$. In fact, the
probability of a positive detection per each light pulse is given
by $P=1-\exp(-\mu \eta)$. Each time a pulse is detected, we have 1
useful trigger and $M$ useless triggers, which fall inside the
detector's dead time. These triggers have to be removed, either by
post-processing or by disabling the trigger before it arrives to
the SPAD. In any case, given $N$ total triggers arriving at the
measuring setup and $k$ positive detections, the percentage of
useful triggers is given by:
\begin{equation}\label{eq:eff}
    P_{\textrm{useful}} = {\ver \frac{N-kM}{N}.}
\end{equation}
The average number of positive detection $\overline{k}$ is found
through a numerical simulation. The result is directly substituted
in Eq.~\eqref{eq:eff} to find the average percentage
$\overline{P}_{\textrm{useful}}$ as function of $\mu$ and $\eta$.
The parameters of the simulation were the following: $\mu = 0.1$,
$\eta = 0.1$, $M = 20$, $N = 8192$, resulting in
$\overline{k}\simeq 68$ and $\overline{P}_{\textrm{useful}}\simeq
{\ver 83.4}$. This last value has been used in
Fig.~\ref{fig:triggers} to plot the dotted curve corresponding to
the theoretical model.
The number of useful triggers per block of cache memory fluctuates
in time. Short-term fluctuations are caused by detector's random
behavior while long-term fluctuation are caused by an imperfect
photon source whose intensity fluctuates. The trigger-disabling
technique {\ver makes} 
the amount of useful triggers {\ver per block of cache memory}
independent of these factors, {\ver as showed in
Fig.~\ref{fig:triggers}, because only useful triggers are present
in that case}.

\section{Pairwise detection and self-blinding}
\label{sec:pair-det-risk-QKD}

The technique described in the previous section can be fruitfully
implemented when two or more SPADs are used at the same time, for
instance in experiments where coincidence detection is important,
or in QKD implementations, where each detector is assigned with a
logical value of one bit, `0' or `1', and with a measurement
basis, $Z$ or $X$. The sequence of the bits will constitute, after
a classical distillation procedure, the secret quantum key used
for secure telecommunications.

For the moment we consider a practical setup with only two
detectors, which we call respectively $D_0$ and $D_1$, assigned
with the value `0' or `1' of the key bit. Such a description well
adapts to what routinely happens in QKD implementations when,
e.g., the BB84 protocol~\cite{BB84} is performed with an active
choice of the basis (see e.g.~\cite{Gobby2004}
and~\cite{Kumar2008_two-way_qkd}). In this case it is crucial for
security that both detectors are active or inactive at the same
time. If one detector is active and the other one is blind, e.g.
during the dead-time following a positive detection event, then a
security risk comes about~\cite{Weier11}.

In the standard scenario of QKD there are two main users,
traditionally called Alice (the transmitter) and Bob (the
receiver), who try to communicate privately over an insecure
quantum channel, plus one eavesdropper, Eve, who aims to steal
information from the channel. Eve can use any means allowed by the
laws of physics to reach her goal.
{\red She can also exploit an imperfection of the setup to her own
advantage. For example in~\cite{Weier11} it is described how to
exploit the detectors' dead-time to hack a particular QKD setup
when the detectors are used in free-running mode. The same
technique, which is a variant of the ``faked-state
attack''~\cite{MH05}, can be adapted to hack also a setup based on
gated-mode detectors, if a narrow temporal selection inside the
gate is effected by the users to further clean their quantum
signal~\cite{Weier11}. This technique, though very powerful,
requires an active intervention by Eve, who has to carefully
design her light pulses and synchronize them with those
transmitted by Alice to Bob. Furthermore, it is not effective
against a QKD setup based on gated-mode detectors in which there
is no temporal selection inside the gate. In that case, the
eavesdropping described in~\cite{Weier11} would cause an abnormal
number of double counts, easily detectable by the legitimate
users.

In the following, we restrict our attention to such a kind of QKD
setup, using gated-mode SPADs with no temporal filter inside the
gate. We describe a totally \textit{passive} action by Eve which
is sufficient to create a security risk for the communication. Our
aim is not to provide an attack which gives Eve 100$\%$ of the
information sent by Alice; we only want to show that some extra
bit could possibly be captured by Eve without the legitimate users
being aware of that.
}

{\red Let us focus then on the typical situation of self-blinding:
at Bob's side one detector is active while the other one is blind.
This can happen in a QKD setup with a trigger rate higher than the
inverse of detector's dead-time, which we define here as
$\tau_{d}$. After the whole quantum communication is completed,
Bob announces on a public channel the addresses corresponding to
his non-vacuum counts; all the remaining addresses are associated
to vacuum counts and are discarded by the users. Eve will register
all the counts announced by Bob, in particular those which
correspond to qubits distant in time less than $\tau_{d}$. All
these bits will be necessarily anti-correlated. In fact, suppose
that at a certain time $t'$ detector $D_0$ fires and Bob registers
a `0'. If a second event occurs at time $t''$ with $|t''-t'|<
\tau_{d}$, it must come necessarily from $D_1$, because $D_0$ is
in its dead-time period. So Bob will see a `1'. If another count
occurs at time $t'''$ with $|t'''-t''|< \tau_{d}$ it must
necessarily be a `0', and so on. The net result will be that with
a small, but non-zero, probability there will be in the final key
groups of bits which are anti-correlated (e.g. 1010101...) rather
than perfectly random. This is already a security breach in the
theory of QKD. In fact, the two requisites for unconditional
secure communication are (i) that the final key should be random
and (ii) that it should be known only to the legitimate
users~\cite{Shannon1949}. So the first requisite here is violated.
}

{\red Now we want to show that also the second requisite of
security can be violated by the passive strategy described above,
i.e. some bits of the final key can be learned by Eve without
Alice and Bob being aware of that. In fact, after the addresses of
the non-vacuum counts have been announced, the standard
description of QKD~\cite{GRT+02} prescribes that Alice and Bob
proceed with two classical procedures known as Error Correction
(EC)~\cite{Brassard1994} and Privacy Amplification
(PA)~\cite{Bennett1995}. EC aims at correcting potential errors
(i.e. different bits) in the users' final keys. During this
procedures a few bits are revealed on public by the users to
localize and finally correct the errors. Eventually the wrong bits
and their positions will be known to Eve too. With PA, the users
are supposed to reduce Eve's knowledge about these bits to nearly
zero. However it can happen that some of these bits fall into a
group which contains anti-correlated bits, e.g. 10101. In that
case it is plain that Eve will immediately know all the members of
the group as soon as she knows a single bit of the group. Hence
the PA will erase Eve's knowledge about the single bit revealed,
but not her knowledge about the remaining bits.
}

{\red The attack just described is entirely passive, as it is} the
result of a bad setup configuration by the users. For this reason
we called it \textit{self-blinding}. As already mentioned,
self-blinding can occur when detector's dead-time is
\textit{higher} than the inverse of the trigger rate:
\begin{equation}\label{basic-cond}
   \tau_{d} > \frac{1}{\omega_{trigger}}.
\end{equation}
The truer this inequality, the easier the self-blinding mechanism.
It is worth noting that the above self-blinding condition,
Eq.~\eqref{basic-cond}, is easy to fulfill, so it could represent
a common mistake when using the SPADs. The reason is that one
usually {\red sets the electronics} for a long dead-time, to
reduce the after-pulses, and at the same time for a high trigger
speed, to increase the final transmission rate.

{\red One solution against self-blinding} is to slow down the
triggering rate of the QKD apparatus, {\red until
Eq.~\eqref{basic-cond} is no more satisfied}, but this
dramatically reduces the total efficiency of the system.
Another possibility is {\red to remove all the bits featuring a
temporal distance less than $\tau_{d}$ by post-processing, but
this can be very time-consuming.
Also, one could resort to alternative descriptions of
QKD~\cite{Koa05,Koa06} in which the bits for EC are encrypted, so
the positions of the errors will be not known to Eve at all. But
this solves only the second flaw, not the first, i.e., the bits of
the final key would still be not purely random.}
{\red Our solution is based on hardware. We simply} apply the
trigger-disabling technique to guarantee that {\red Bob's}
detectors are {\red both} active or {\red both inactive} at the
same time.

%
In the next Section we describe how this can be put into practice
and present the experimental results obtained with this technique.

\section{Experimental pairwise detection with trigger-disabling}
\label{sec:exp-coinc}

In order to disable the triggering clock in case of a pairwise
detector configuration we used a modified version of the circuit
shown in Fig.~\ref{fig_scheme}.
We connected both detectors outputs to the clock input port C of
FFD through an OR gate, and fed their trigger
input with the same signal coming from CLOCK OUTPUT.

To show the effectiveness of the trigger-disabling technique we
applied the electronics to a real setup composed by two SPADs, one
id201, as before, and one id200. Both detectors are triggered at
4~MHz. We notice that the id201 detector is one of the receiving
units in the ``All-Vienna'' QKD system~\cite{Treiber_AllVienna},
used in two recent quantum networks~\cite{SECOQC_network,
Sasaki2011_QKDnetwork}. According to what reported in
Ref.~\cite{Treiber_AllVienna}, the system's average trigger rate
was of 415~KHz. Hence Eq.~\eqref{basic-cond} entails that
self-blinding becomes important for that setup if a dead-time
longer than 2.4~$\mu$s is used.
On the other side, the id200 detector was used in the QKD system
described in~\cite{Kumar2008_two-way_qkd}, and in the setup for
the asymmetric feedback adopted
in~\cite{Lucamarini2010_phase_drift_compensation}. In
Ref.~\cite{Kumar2008_two-way_qkd} the trigger rate was 2.5~MHz and
the dead-time 10~$\mu$s, so the self-blinding could have had an
influence in that case. However, it should be noted that the QKD
protocol used was not the BB84 but rather the
LM05~\cite{Lucamarini2005a}, in which the two detectors are not
directly associated to the logical bit value. Hence the above
analysis is not directly applicable to this case.


In the present paper's experiment we studied two figures of the
apparatus, i.e., the frequency of coincidence counts and the
randomness of the final key. In both cases we employed a light
with intensity higher than the single-photon level, to simulate
the detectors' response under the Eve attack described above. In
the previous experiment the value of $\mu$ was about $0.1$. Now it
is $\mu \simeq 4$.

In order to see the frequency of coincidence counts, we collected
more than \(10^7\) events. We registered \(13.5\%\) of
coincidences with trigger disabling ON and only \(3.0\%\) with
trigger disabling OFF. This confirms that { our technique reduces
the probability of a self-blinding, because the two detectors are
always operational together.} Notice that the coincidence rate is
not 100\% even when trigger disabling is ON, because the light is
not intense enough. A much more intense light would cause 100\%
coincidence counts if the trigger disabling is ON, but could also
damage the very sensitive SPAD.

\begin{figure}[!h]
\begin{center}
    \includegraphics[width=7cm]{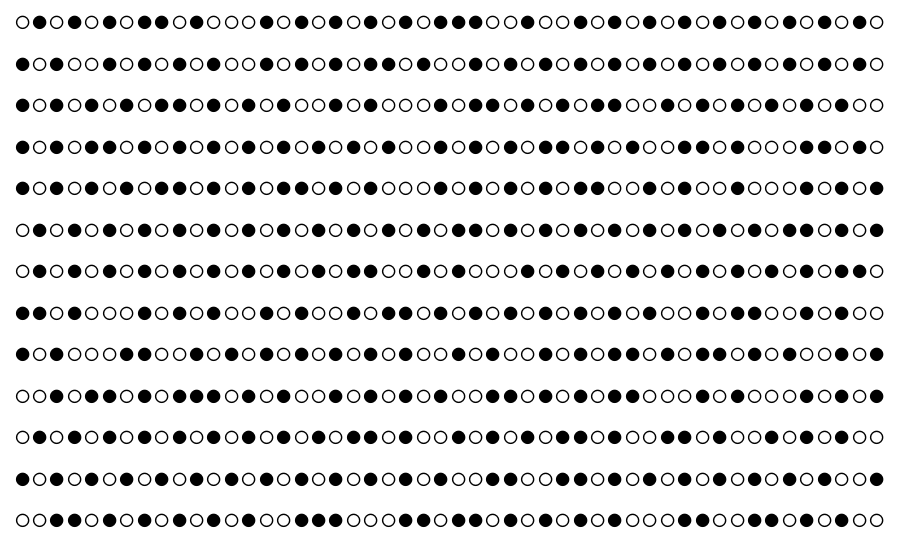}\\
    \vspace{0.5cm}
    \includegraphics[width=7cm]{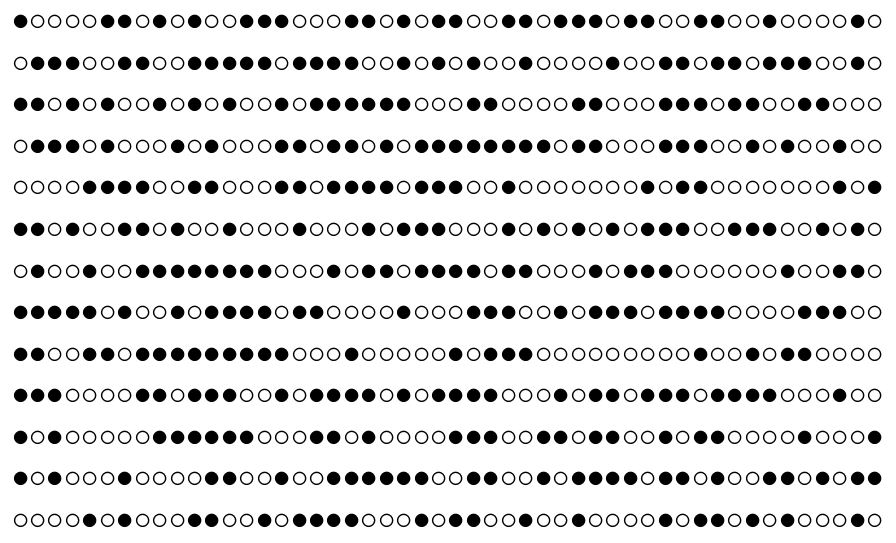}\\
    \caption{Sequences of events corresponding to detector $D_0$ firing and $D_1$ not firing
    (empty circles) or $D_1$ firing and $D_0$ not firing (filled circles).
    \textit{Top panel}: a sequence of clearly
    non-random events, obtained when the trigger-disabling technique was OFF. The alternate
    occurrence of empty and filled circles is apparent. \textit{Bottom panel}: a sequence
    of events looking like random, obtained when the trigger-disabling technique is ON.
    The occurrence of relatively long sequences of empty and filled dots together with
    short sequences witnesses the random-like nature of the string. The events with both
    detectors firing have been removed for explicative purposes. For the measurement we employed
    a light with average photon-number $\mu \simeq 4$.}
    \label{fig:rand-nonrand}
\end{center}
\end{figure}

On a second step we monitored the randomness of the strings
obtained from detectors when trigger disabling was either ON or
OFF. Specifically we assigned a click from detector $D_0$ ($D_1$)
and no click from detector $D_1$ ($D_0$) with the value `0' (`1'),
and collected all the values so to form two strings, one
corresponding to ON and one to OFF. Two short sequences extracted
from such strings are reported in Fig.~\ref{fig:rand-nonrand};
the empty (filled) circles correspond to the value `0' (`1'). The
difference is quite apparent. When trigger disabling is OFF (upper
string), the sequence is clearly non random, since almost always a
single 0 is followed by a single 1, and viceversa. On the
contrary, when trigger disabling is ON (lower sequence), the
occurrence of a 0 or a 1 is much less foreseeable. Also, the
presence of relatively long sub-strings filled with all 0's or
1's, is an additional evidence of a random behavior. Note that
while non-randomness can be easily demonstrated, for true
randomness of finite sequences is not that easy since it does not
exist a decisive test in this respect. Nevertheless we performed
the DIEHARDER statistical tests~\cite{dieharder} on our
strings, most of which were passed by the ON string and none was
passed by the OFF string.

{\red It could be argued that our choice $\mu \simeq 4$ is
somewhat unusual in a QKD setup, where one has more often $\mu
\simeq 0.1$. Our choice is motivated by the purpose of showing
clearly in the experiment the effect of self-blinding. When $\mu$
is smaller, the consequences of self-blinding are smaller too, but
they do not disappear as there still remains a non zero
probability to find blocks in the final key containing
anti-correlated, non random, bits.}

\section{Conclusion}
\label{sec:conclusion}

We introduced a FPGA-driven technique to run a single-photon
avalanche diode at its maximum trigger rate, regardless of
dead-time limitation. While in standard situations some care
should be paid when the trigger rate is higher than the inverse of
dead-time, with our technique this hindrance is removed by a
trigger-disabling loop which stops the trigger to and from the
SPAD until it has properly recovered from the dead-time period.
The presented circuit can also serve as a simple dead-time
generator for single-photon detectors which are produced without
their own generator.

{ Because of the absence of any futile counts, our technique is
especially useful in case of implementation in embedded QKD
systems as in this case one needs to record a huge amount of
results in a limited local memory. Moreover the technique allows
to remove completely the post-processing related to futile zeroes
discarding, what results in shorter acquisition time.}

We provided an experimental evidence of our technique applied to a
single SPAD, by reporting the trigger distribution in the
acquisition system's cache memory. Additionally, we adapted the
acquisition system to a pair of detectors used in coincidence and
found an evident advantage of the trigger-disabling technique over
a standard trigger usage in terms of distribution of the
coincidence counts and in terms of randomness of the final string
distilled from detectors' counts. { Moreover, it removes at the
roots the possibility of an external attack based on
``self-blinding'', because it forces the detectors in the
receiving unit to be active or inactive altogether, at the same
time.}

The proposed technique can be easily implemented in many different
setup by just adding a piece of electronics in front of a standard
SPAD, thus making the detector response time as short as possible.
This paves the way to a future straightforward implementation of
this technique in integrated commercial QKD apparatuses.

\section*{Acknowledgment}

We thank L. Widmer for his feedback on IDQ detectors and R. Kumar
for useful discussions. M.L. is supported by the
$5\tcperthousand$~grant C.F.~81001910439.


\begin{thebibliography}{99}

\bibitem{note1} Here and throughout the paper the reader can think to a
single-photon-attenuated light as to a coherent state with mean
photon-number $\mu=0.1$, as the one coming, e.g., from an
attenuated pulsed diode-laser. This is a standard intensity in
several QKD systems.

\bibitem{ZTC+07} F.~Zappa, S.~Tisa, S.~Cova, P.~Maccagnani, R.~Saletti, R.~Roncella,
F.~Baronti, D.~Bonaccini Calia, A.~Silber, G.~Bonanno and M.~Belluso,
J.~Mod. Opt. \textbf{54}, 163 (2007).

\bibitem{ELZ+10} P.~Eraerds, M.~Legre, J.~Zhang, H.~Zbinden,
N.~Gisin, IEEE J.~Light. Techn. \textbf{28}, 952 (2010).

\bibitem{BB84} C.~H.~Bennett and G.~Brassard, Proc. IEEE Int. Conf. Comp., Sys.
and Sign. Process., p. 175 (1984).

\bibitem{GRT+02} N.~Gisin, G.~Ribordy, W.~Tittel, and H.~Zbinden, Rev. Mod. Phys. \textbf{74}, 145 (2002).

\bibitem{Scarani2009} V.~Scarani, H.~Bechmann-Pasquinucci, N.~J.~Cerf, M.~Dusek, and N.~L\"{u}tkenhaus, and M.~Peev, Rev. Mod. Phys. \textbf{81}, 1301 (2009).

\bibitem{idq} www.idquantique.com.

\bibitem{Namekata2006} N.~Namekata, S.~Sasamori and S.~Inoue, Opt. Express \textbf{14}, 10043 (2006).

\bibitem{Yuan2007} Z.~L.~Yuan, B.~E.~Kardynal, A.~W.~Sharpe, and A.~J.~Shields, Appl. Phys. Lett. \textbf{91}, 041114 (2007).

\bibitem{Dixon2009} A.~R.~Dixon, J.~F.~Dynes, Z.~L.~Yuan, A.~W.~Sharpe, A.~J.~Bennett and A.~J.~Shields, Appl. Phys. Lett. \textbf{94}, 231113 (2009).

\bibitem{Zhang2009} J. Zhang, R.~Thew, C.~Barreiro and H.~Zbinden, Appl. Phys. Lett. \textbf{95}, 091103 (2009).

\bibitem{Namekata2009} N.~Namekata, S.~Adachi, and S.~Inoue, Opt. Expr. \textbf{17}, 6275 (2009).

{\red
%
\bibitem{MBH04} V. Makarov, A. Brylevski, and D. R. Hjelme, Appl. Opt. \textbf{43}, 4385
(2004).
%
}


\bibitem{note2} { We directly measured this value on id200. A similar value can be found in the datasheets of the
id220~\cite{idq}.}


{\red
%
\bibitem{Weier11} H. Weier, H. Krauss, M. Rau, M. F\"{u}rst, S. Nauerth, and H. Weinfurter, New J. Phys. \textbf{13}, 073024 (2011).
%
}

\bibitem{Mak09} V.~Makarov, New J.~Phys. \textbf{11}, 065003 (2009).

\bibitem{XilinxUG:Clocking} Xilinx User Guide, p.~42, \emph{Spartan-6 FPGA Clocking
Resources}, v1.4.\hskip 1em plus 0.5em minus 0.4em\relax Xilinx,
2010.

\bibitem{note4} In our setup the post-processing was performed on a
MicroBlaze architecture and took 99$\%$ of the total acquisition
time. This means that of 12 hours only about 7 minutes were really
necessary for the experiment, the remaining time being devoted to
post-processing. Even though post-processing can be certainly done
in a more efficient way, it cannot compete with the zero time
assured by our hardware solution.

\bibitem{Gobby2004} C. Gobby, Z. L. Yuan, and A. J. Shields, Appl. Phys. Lett. \textbf{84}, 3762
(2004).

\bibitem{Kumar2008_two-way_qkd} R. Kumar, M.~Lucamarini, G.~Di~Giuseppe, R.~Natali, G.~Mancini, and P.~Tombesi
Phys. Rev. A \textbf{77}, 022304 (2008).

\bibitem{MH05} V. Makarov and D. R. Hjelme, J. Mod. Opt. \textbf{52},
691 (2005).

{\red
%
\bibitem{Shannon1949} C. E. Shannon, Bell Syst. Techn. Journ. \textbf{28}, 656
(1949).
%
}

\bibitem{Brassard1994} G.~Brassard and L.~Salvail, Lecture Notes in Computer Science,
Advances in Cryptology -- EUROCRYPT '93 \textbf{765}, 410 (1994).

\bibitem{Bennett1995} C. H. Bennett, G. Brassard, C. Cr\'{e}peau, and U. M. Maurer,
IEEE Trans. Inf. Th. \textbf{41}, 1915 (1995).

{\red

\bibitem{Koa05} M. Koashi, e-print quant-ph/05051080v1.

\bibitem{Koa06} M. Koashi, J.of Phys. Conference Series, Vol. \textbf{36}, 98 (2006).

}



\bibitem{GLL+11} I.~Gerhardt, Q.~Liu, A.~Lamas-Linares, J.~Skaar, C.~Kurtsiefer,
and V.~Makarov, Nat. Comm. \textbf{2}, 349 (2011).

\bibitem{Treiber_AllVienna} A.~Treiber, A.~Poppe, M.~Hentschel, D.~Ferrini, T.~Lor{\ver \"{u}}nser, E.~Querasser, T.~Matyus, H.~H{\ver \"{u}}bel, and A.~Zeilinger, New J.~Phys.  \textbf{11(4)}, 045013 (2009).

\bibitem{SECOQC_network} M. Peev, \textit{\underline{et al.}}, New J. Phys.  \textbf{11(7)}, 075001 (2009).

\bibitem{Sasaki2011_QKDnetwork} M. Sasaki, \textit{\underline{et al.}}, Opt. Expr. \textbf{19}, 10387 (2011).

\bibitem{Lucamarini2010_phase_drift_compensation} M.~Lucamarini, R.~Kumar, G.~Di~Giuseppe, D.~Vitali, and
P.~Tombesi, Phys. Rev. Lett. \textbf{105}, 140504 (2010).

\bibitem{Lucamarini2005a} M.~Lucamarini and S.~Mancini, Phys. Rev. Lett. \textbf{94}, 140501
(2005).

\bibitem{dieharder} A~Random Number Test Suite, 2011, URL http://www.phy.duke.edu/\texttildelow rgb/General/dieharder.php, version 3.31.0 by Robert~G. Brown







\end{thebibliography}
\end{document}